\documentclass[aps,prl,amsmath,amssymb,twocolumn,superscriptaddress,a4paper]{revtex4-1}
\usepackage{graphicx}
\usepackage{dcolumn}
\usepackage{bm}
\usepackage[breaklinks]{hyperref}
\usepackage{color}
\usepackage[usenames,dvipsnames,svgnames,table]{xcolor}
\usepackage{amsmath}
\usepackage{colordvi}
\usepackage{soul}

\begin{document}
\title{Microscopic mechanism for the shear-thickening of non-Brownian suspensions}
\author{Nicolas Fernandez}
\affiliation{Laboratory for Surface Science and Technology,  Department of Materials, ETH Zurich, Switzerland}
\author{Roman Mani}
\affiliation{Computational Physics for Engineering Materials, Department of Civil, Environmental and Geomatic Engineering, ETH Zurich, Switzerland}
\author{David Rinaldi}
\affiliation{Lafarge LCR, Saint Quentin-Fallavier, France}
\author{Dirk Kadau}
\affiliation{Computational Physics for Engineering Materials, Department of Civil, Environmental and Geomatic Engineering, ETH Zurich, Switzerland}
\author{Martin Mosquet}
\affiliation{Lafarge LCR, Saint Quentin-Fallavier, France}
\author{H\'{e}l\`{e}ne Lombois-Burger}
\affiliation{Lafarge LCR, Saint Quentin-Fallavier, France}
\author{Juliette Cayer-Barrioz}
\affiliation{Laboratoire de Tribologie et Dynamique des Syst\`{e}mes - UMR 5513 CNRS, \'{E}cole Centrale de Lyon, France}
\author{Hans J. Herrmann}
\affiliation{Computational Physics for Engineering Materials, Department of Civil, Environmental and Geomatic Engineering, ETH Zurich, Switzerland}
\author{Nicholas D. Spencer}
\affiliation{Laboratory for Surface Science and Technology, Department of Materials, ETH Zurich, Switzerland}
\author{Lucio Isa }
\email[Corresponding author: ]{lucio.isa@mat.ethz.ch}
\affiliation{Laboratory for Surface Science and Technology,  Department of Materials, ETH Zurich, Switzerland}

\begin{abstract}

{We propose a simple model,  supported by contact-dynamics simulations as well as rheology and friction measurements, that links the transition from continuous to discontinuous shear-thickening in dense granular pastes to distinct lubrication regimes  in the particle contacts.  We identify a local Sommerfeld number that determines the transition from Newtonian to shear-thickening flows, and then show that the suspension's volume fraction and the boundary lubrication friction coefficient control the nature of the shear-thickening transition, both in simulations and experiments.
}
\end{abstract}
\maketitle
{
Flow non-linearities attract fundamental interest and have major consequences in a host of practical applications \cite{Larson:1999ud,Coussot:1999wg}. In particular, shear-thickening (ST), a viscosity increase from a constant value (Newtonian flow-Nw) upon increasing shear stress (or rate) at high volume fraction $\phi$, can lead to large-scale processing problems of dense pastes, including cement slurries  \cite{Barnes:1989vr}. Several approaches have been proposed to describe the microscopic origin of shear-thickening \cite{Andrade:1949wo,Bagnold:1954ek,Bossis:1985wr,Hoffman:1972wt}. The most common explanation invokes the formation of "hydroclusters", which are responsible for the observed continuous viscosity increase \cite{Bossis:1985wr,Boersma:1990wn,BRADY:2009wj} and which have been observed for Brownian suspensions of moderate volume fractions \cite{BERGENHOLTZ:2002if,Cheng:2011do}. However, this description no longer holds for bigger particles and denser pastes, where contact networks can develop and transmit positive normal stresses \cite{Cates:1998td}. Moreover, the link between hydroclusters and CST for non-Brownian suspensions is still a matter of debate \cite{Morris:2009kg}. Additionally, dense, non-Brownian suspensions can also show sudden viscosity divergence under flow \cite{Fall:2008wh,Isa:2009kv,Brown:2009wj,Brown:2010ex} with catastrophic effects, such as pumping failures. In contrast to a continuous viscosity increase at any applied rate, defined as continuous shear-thickening (CST), the appearance of an upper limit of the shear rate defines discontinuous shear-thickening (DST). This CST to DST transition is observed when the volume fraction of the flowing suspension is increased above a critical value, which depends on the system properties, e.g. polydispersity or shape, and on the flow geometry \cite{Barnes:1989vr,Brown:2010wt}. An explanation for its microscopic origin is still lacking \cite{Coussot:2009ua}. Moreover, experiments have demonstrated that the features of the viscosity increase (slope, critical stress) can be controlled by tuning particle surface properties such as roughness \cite{Lootens:2005wi} and/or by adsorbing polymers \cite{Toussaint:2009wa,LomboisBurger:2008td}. These findings suggest that inter-particle contacts play a crucial role in the macroscopic flow at high volume fractions. A more precise description of these contacts is therefore essential to interpret the rheological behavior.

In this paper, we present a unified theoretical framework, supported by both numerical simulations and experimental data, which describes the three flow regimes of rough, frictional, non-Brownian particle suspensions (Nw,CST,DST) and links the Nw-ST (in terms of shear) and the CST-DST transitions (in terms of volume fraction) to the local friction. Our microscopic particle-contact based description, as opposed to macroscopic scaling, explains both the occurrence of DST and recovers Bagnold's analysis \cite{Bagnold:1954ek} for CST, respectively above and below a critical volume fraction.

$\newline$

The lubricated contact between two solid surfaces has been widely studied in the past \cite{Stachowiak:2005va}. It is now commonly accepted that different lubrication regimes occur as a function of a characteristic number, the Sommerfeld number $s$. For two identical spheres, $s=\eta_{f} \textmd{v}R_p/N$, where $\eta_{f}$ is the fluid viscosity, $\textmd{v}$ is the sliding speed between the two solid surfaces, $R_p$ is the radius of the spheres and $N$ is the normal load. At high $s$ ("hydrodynamic regime"-HD), a fluid film fully separates the two sliding surfaces and the friction coefficient $\mu$ depends on $s$. For low $s$, below a critical value $s_c$, the lubrication film breaks down and contacts between the microscopic asperities on each surface support most of the load. This "boundary lubrication" regime (BL) exhibits friction coefficients that only very weakly depend on $s$. For intermediate values of $s$ the system is in a "mixed regime", where the sharpness of the transition depends on the system properties (e.g. contact roughness, rheology of the fluid. See Fig.\ref{Stribeckandcurves}a) \cite{Stachowiak:2005va}.
$\newline$

\begin{figure}
\includegraphics[width=0.45\textwidth]{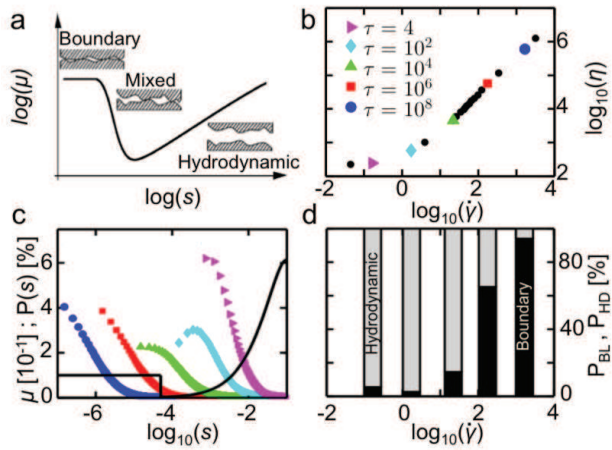}\\
\caption{(color online) a) Schematic Stribeck curve. Evolution of the friction coefficient, $\mu$, versus the Sommerfeld number, $s$, for a lubricated contact.  b) Apparent viscosity, $\eta$, versus the shear rate, $\dot\gamma$, from the numerical simulations. c) Numerical simulations friction law (black line) and probability distributions of $s$, $P(s)$, for all contacts at several shear stresses as defined in b. d) Frequencies of BL contacts, $P_{BL}$, and HD contacts, $P_{HD} = 1-P_{BL} $, as a function of $\dot\gamma$ for the stresses defined in b. The simulations data in b-c-d have $\phi= 0.58$, $\mu_0 = 0.1$ and $s_c = 5 \times 10^{-5}$.}\label{Stribeckandcurves}
\end{figure}

Both experiments and models show that Nw flow is stable below a critical shear rate $\dot\gamma_c$ where the contacts between particles are HD lubricated.
On the other hand, a particle-contact-dominated flow requires, by definition, that $s<s_c$ and it is equivalent to a dense dry granular flow (i.e. no suspending fluid lubrication effects). Dense granular flows follow a quadratic scaling of the normal and shear stress $P$ and $\tau$ with the shear rate $\dot\gamma$ (Bagnold scaling) through a volume-fraction-dependent factor \cite{Bagnold:1954ek}; this implies that the apparent viscosity rises linearly with $\dot\gamma$ and that the system shear thickens continuously (see Fig.\ref{Stribeckandcurves}b). This scaling can be expressed in terms of a dimensionless parameter, the inertial number $I= \dot\gamma R_{p} \sqrt{\frac{\rho_{p}}{P}}$, only depending on $\phi$ and $\mu$ for rigid particles with density $\rho_{p}$ \cite{Forterre:2008vk}.

Given the definition of $s$, this leads to $\ s \propto \eta_{f} I^{2}/\dot\gamma \rho_{p} R_{p}^2$. This Bagnold (CST) regime is possible as long as $\dot\gamma$ is larger than $\dot\gamma_c \propto \eta_{f} I^{2}/s_c \rho_{p} R_{p}^2$, showing the link between $\dot\gamma_c$ and $s_c$ when particle contacts dominate. This transition was partially proposed, with macroscopic arguments, by Bagnold \cite{Bagnold:1954ek,Trulsson:2012dd,Boyer:2011vz}. Nevertheless, our microscopic analysis also accounts for volume fraction effetcs.
$\newline$

In our model, the existence of two lubrication mechanisms (boundary and hydrodynamic) implies two different jamming volume fractions $\phi_{max}$, above which flow is not possible. If the system is hydrodynamically lubricated, the jamming volume fraction $\phi^{HD}_{max}$ is at random close packing $\phi_{RCP}$, regardless of the boundary friction coefficient \cite{Roussel:2010uf}. Conversely, when the system is in a boundary-lubricated Bagnold regime, the jamming volume fraction $\phi^{BL}_{max}$ decreases with $\mu$ \cite{Silbert:2010by,Ciamarra:2011ju}. Both $\phi^{HD}_{max}$ and $\phi^{BL}_{max}$ are independent of $\dot\gamma$ for non-Brownian particles. It follows that $ \phi_{RCP}=\phi^{HD}_{max} \geq \phi^{BL}_{max}(\mu)$. When $\phi \leq \phi^{BL}_{max} \leq \phi^{HD}_{max}$, the transition from hydrodynamic to boundary-dominated flow is possible and the suspension exhibits CST, as reported above and predicted by Bagnold. When $\phi^{BL}_{max} < \phi \leq \phi^{HD}_{max}$, the transition to a Bagnold regime is forbidden, and the shear rate cannot exceed $\dot\gamma_c$: the system undergoes DST. As a consequence, $\phi^{BL}_{max}$ is the critical volume fraction for DST and therefore it can be tuned by changing the particle friction coefficient. Both numerical simulations and experiments fully and independently support our model.
$\newline$

In concentrated systems most of the dissipation arises from particles that are in, or close to, contact and not from Stokesian drag \cite{Frankel:1967ey,Trulsson:2012dd}. This motivates using Contact Dynamics \cite{Moreau:1994ui,Jean:1992ts,Brendel:2004tb,Mani2012,Kadau2009} to simulate dense suspensions of hard, spherical, frictional particles using a simplified Stribeck curve (no mixed regime) as friction law (see Fig.\ref{Stribeckandcurves}c and Eq.\ref{friction_law}). Only one dissipative mechanism, either BL or HD, is taken into account in each contact. This constitutes the simplest physical description of a lubricated contact.

The boundary lubrication between two rough particles is described using Amontons-Coulomb friction, i.e. the coefficient of friction $\mu_0$ being independent of the load, the speed and the apparent contact area \cite{Stachowiak:2005va}.

In the HD regime, the hydrodynamic interactions between two neighboring particles are long-lived and can be described by standard, low-Reynolds-number fluid mechanics with a lubrication hypothesis \cite{BarthesBiesel:2012wx}, from which a friction coefficient can be calculated as a function of the Sommerfeld number $\mu = 2 \pi s ln(\frac{5}{6 \pi s} )\label{eq:logviscous}$ (see Supplemental Material for full derivation). The lubrication hypothesis breaks down when the particles are too far apart (i.e when $s$ is large) and therefore we consider only a range of $\dot\gamma$ where $s$ of almost all the contacts is smaller than a limit value, $s_{lim}=10^{-1}$.

The friction law used for the simulations is then:
\begin {equation}
\label{friction_law}
\mu(s)= \left\{
\begin{array}{ll}
\mu_0 & \mbox{if } s < s_c \\
2 \pi s\ln(\frac{5}{6 \pi s}) & \mbox{if } s_c< s < s_{lim}\\
\end{array}
\right.
\end{equation}

In our Contact Dynamics simulations the normal forces are calculated based on perfect volume exclusion, using zero normal restitution coefficient, and we simulate stress-controlled ($\tau$) simple shear between moving and fixed rough walls (obtained by randomly glued particles) at a constant volume fraction \cite{Latzel2003,Pena2009}. The rectangular simulation box dimensions are $(L_x, L_y, L_z)=(25R,10R,27R)$, where $L_z$ is the distance between the two walls and $R$ the radius of the largest particle in the simulations. We use periodic boundary conditions in both $x$ and $y$ directions. The presence of hard confinement mimics experimental conditions, and simulations with Lee-Edwards boundary conditions that are periodic in the three directions show the same qualitative behavior (see Supplemental Material). The particle radii are uniformly distributed between $0.8R$ and $R$ to prevent crystallization. When fixing $\phi$, $\mu_0$, $R$, $\rho_p$ and $s_c$, the physics of the system is characterized by a single dimensionless number: $\lambda=\sqrt{\tau\rho}R/\eta_f$. $\lambda$ can be understood as the ratio between the microscopic time scale of the lubricating fluid, $\eta_f / \tau$, and of the granular medium, $R\sqrt{\rho/\tau}$ \cite{Forterre:2008vk}. Increasing the shear stress $\tau$ quadratically is equivalent to decreasing $\eta_f$ linearly. In our simulations, we vary $\eta_f$ and keep $\tau$ fixed. After the system has reached its steady state with a linear velocity profile (no shear bands), we measure the time averaged velocity of the moving wall $\langle v_{wall}\rangle$, thus $\dot\gamma = \langle v_{wall}\rangle/L_z$ and the apparent viscosity of the suspension $\eta$ is given by $\tau/\dot\gamma$. The quantities $\dot\gamma$, $ \tau$ and $\eta$ are measured in units of $\eta_f/\rho_p R^2$, $\eta_f^2/\rho_p R^2$ and $\eta_f$ (see Supplemental Material for details).

The simulations (see Figs.\ref{Stribeckandcurves}b and \ref{fric_comp}) reproduce a transition between a Newtonian regime at low shear rates (independent of $\mu_0$ and dominated by HD-lubricated contacts) to a ST regime with increasing $\dot\gamma$, for which boundary lubricated contacts are dominating. In the absence of hydrodynamics in the friction law, such a transition is lost (see Supplemental Material). Indeed, in Fig.\ref{Stribeckandcurves}c for increasing applied stress, the distributions of $s$ in all the particle contacts shift toward the BL regime in the Stribeck curve. In our simulations, the system shear thickens when at least $\approx20\%$ of the contacts are below $s_c$. In Fig.\ref{Stribeckandcurves}d, the percentage of particles in BL and HD contacts is plotted against $\dot\gamma$ for the stresses defined in Fig.\ref{Stribeckandcurves}b. For low $\mu_0$, this ST regime is continuous and fits with a Bagnoldian scaling  ($\eta \propto  \dot \gamma$). Here, the viscosity increases with $\mu_0$, as in a dry granular medium \cite{Forterre:2008vk}. This scenario changes as $\mu_0$ goes beyond a critical value, here 0.35 for $\phi =0.58$ (Fig.\ref{fric_comp}). Then, the system cannot be sheared above a critical shear rate for any shear stress: the system shear-thickens discontinuously.

\begin{figure}[htbp]
\begin{center}
\includegraphics[width=0.45\textwidth]{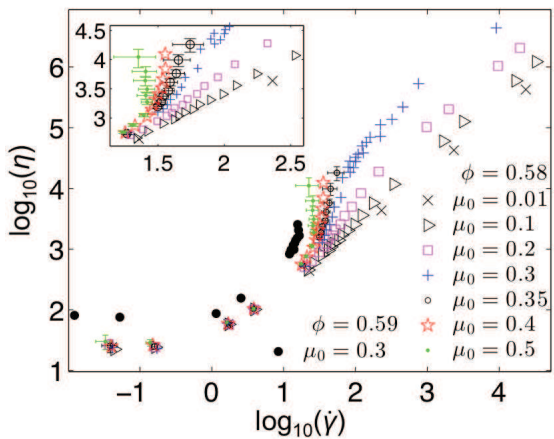}
\caption{(color online)Apparent viscosity versus $\dot\gamma$ for different $\mu_0$ and $s_c=5\cdot 10^{-5}$ in simulations. In the Newtonian regime, the viscosity does not depend on $\mu_0$ but on $\phi$.  At $\phi=0.58$, for $\mu_0 \leq 0.3$, the system experiences CST, where the viscosity depends on the friction coefficient. For $\mu_0 \geq 0.35$, the system jams at sufficiently large $\dot\gamma$. Data points for $\phi=0.59$ show DST for $\mu_0 =0.3$. Inset: Zoom of the transition zone.}
\label{fric_comp}
\end{center}
\end{figure}

The transition from CST to DST does not only occur when increasing $\mu_0$ but also when increasing $\phi$: the system experiences CST at $\phi=0.58$, $\mu_0=0.3$ but experiences DST for $\phi=0.59$ and the same $\mu_0$ (see Fig\ref{fric_comp}). Moreover, as predicted in our theoretical model, the CST-DST transition occurs when $\phi$ is increased above a $\phi_{max}^{BL}(\mu_0$$=$$0.3)$,  compatible with \cite{Silbert:2010by}.

In brief, the numerical simulations confirm that our theoretical framework sets the sufficient conditions to explain Nw-ST and CST-DST transitions.
$\newline$

Our model is also independently supported by experiments where the link between local friction and macroscopic rheology is established using quartz surfaces. We first show experimentally that the volume fraction of the CST-DST transition is indeed $\phi^{BL}_{max}$ and then that it can be tuned by modifying $\mu_{0}$. This is demonstrated by using four different comb polymers, i.e. poly(methacrylic acid) (PMAA) grafted with poly(ethylene glycol) (PEG) side chains, which are dissolved in a $Ca(OH)_2$ saturated aqueous buffer solution with 20 mmol/L $K_2SO_4$. The co-polymers were synthesized by radical polymerization in water according to \cite{Rinaldi:2009tc,Jones:2005dl}. Their specifications, obtained from aqueous gel permeation chromatography (GPC) are (backbone size in kDa, number of carboxylic acids per side chain and side chain size in kDa): Polymer A: PMAA(4.3)-g(4)-PEG(2), Polymer B: PMAA(3.4)-g(2.3)-PEG(2), Polymer C: PMAA(4.3)-g(4)-PEG(0.5) and Polymer D: PMAA(5)-g(1.5)-PEG(2). Once in the buffer solution, these comb polymers are readily adsorbed onto a negatively charged surface, such as quartz, by calcium-ion bridging, and create a stable and highly solvated PEG coating on the solid surface \cite{Flatt:2009ja} that is known to modify the BL coefficient of friction \cite{Lee:2008cb}. The conclusions of the experiments are not dependent on the choice of system, which is a model material for industrial applications (e.g. cement slurry), for which the friction coefficient can be easily tuned.
$\newline$

The  rheological analysis was performed on suspensions of ground quartz (Silbelco France C400, $D_{50} =12 \mu m$) with $\Phi$ between 0.47 and 0.57 in the alkali polymer solutions (see Supplemental Material for details). We initially measured $\phi^{BL}_{max}$ via compressive rheology by high-speed centrifugation (acceleration $\approx 2000g$) of a fairly low concentration suspension ($\phi = 0.47$) in a 10mL measuring flask and calculating the average sediment volume fraction for the various polymers. During sedimentation at high speed, particles come into contact and jam, producing a looser sediment compared to frictionless objects. After 20 minutes of centrifugation, no further evolution is observed and we measured $\phi^{BL}_{max}(A) = 0.578$, $\phi^{BL}_{max}(B)=0.560$, $\phi^{BL}_{max}(C)=0.555$, $\phi^{BL}_{max}(D)=0.545$ (see Fig.\ref{rheo}a).

\begin{figure}[!tbh]
\includegraphics[width=0.45\textwidth]{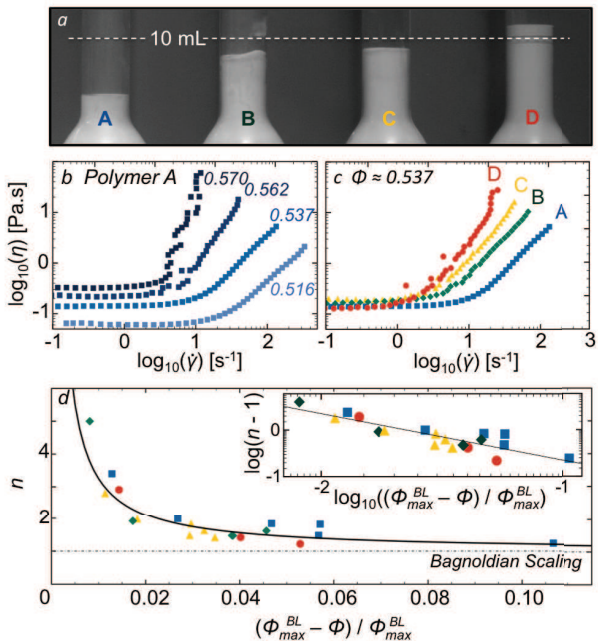}\\
\caption{(color online)
a) Sediment heights for the different polymers after centrifugation.
b) Viscosity vs shear rate with adsorbed polymer A for various $\phi$ of quartz microparticle suspensions.
c) Viscosity vs shear rate for the four adsorbed polymers at analogous $\phi$ ($\phi(A)=0.537$, $\phi(B)=0.537$, $\phi(C)=0.538$, $\phi(D)=0.535$).
d) Oswald-De Waele exponent $n$ vs the reduced volume fraction (same symbols as in c). Inset: Same data in log-log plot. The solid line is a power-law fit for $(n-1)$ vs reduced volume fraction.}
\label{rheo}
\end{figure}

The CST-DST transition was then measured by shear rheometry in a helicoidal paddle geometry (Anton Paar 301 rheometer, see \cite{Toussaint:2009wa} Fig.4 for geometry description) with a descending logarithmic stress ramp after pre-shear (from $700$ to $0.01$ $Pa$ in 100s). The viscosity curves are divided into two main parts: at low shear rate, the fluid shows a Newtonian behavior with a viscosity that depends on volume fraction \cite{Krieger:1959cs} (Fig.\ref{rheo}b) but not on the polymer coating (Fig.\ref{rheo}c). For high shear rates, the fluid shear-thickens. At moderate volume fractions, the system undergoes CST with $\tau \propto \dot\gamma^{2}$ (Bagnoldian regime) as observed by \cite{Fall:2010hu}, while for the higher volume fractions in our experiment, the abruptness of ST increases quickly at a critical $\Phi$ (see Fig.\ref{rheo}b for Polymer A). Above this threshold, the suspensions display DST. In order to quantify this critical volume fraction, the flow curves for the various $\phi$ in the ST regime are fitted by an Oswald-De Waele power law: $\eta\propto \dot\gamma^{n}$. In Fig\ref{rheo}d, we show that $n (\phi)$ diverges exactly at the polymer-dependent $\phi^{BL}_{max}$ that we measured independently by centrifugation, as predicted by our model. Moreover, the data from the different polymer coatings collapse onto a single master curve as a function of a reduced volume fraction $(\phi_{max}^{BL}-\phi)/\phi^{BL}_{max}$ that does not depend on surface properties. A similar collapse  was observed for particles of different shapes \cite{Brown:2011uca}.
$\newline$

To complete our analysis we measured the BL friction coefficients $\mu_0$ between a polished rose quartz stone surface (Cristaux Suisses, Switzerland) and a 2 mm diameter borosilicate sphere (Sigma-Aldrich, USA) coated with the four different polymers, using a nanotribometer (CSM instruments, Switzerland). The contact was immersed into a drop of polymer solution. The experiments were realized in an $N_2$ atmosphere at sliding velocities between $10^{-5}$ and $10^{-3}$ m/s (see Supplemental Material for a protocol). The measured values of $\mu_0$ reported Fig.\ref{PhiVsMu}in are speed independent, as expected in the BL regime. The differences in the friction for the different polymers have been previously ascribed to a variation of the PEG unit density on the surfaces \cite{Perry:2009um}, stemming from an equilibrium between entropic side chain repulsion and backbone/surface electrostatic attraction (through calcium bridging).

Finally, Fig.\ref{PhiVsMu} shows the direct correlation between the BL coefficients of friction and the measured maximum volume fraction $\phi^{BL}_{max}$ that separates CST and DST, as included in our model.  $\phi^{BL}_{max}$ is a decreasing function of the particle friction coefficient in the boundary regime, as predicted by simulations \cite{Silbert:2010by,Ciamarra:2011ju}.

\begin{figure}
\includegraphics[width=0.45\textwidth]{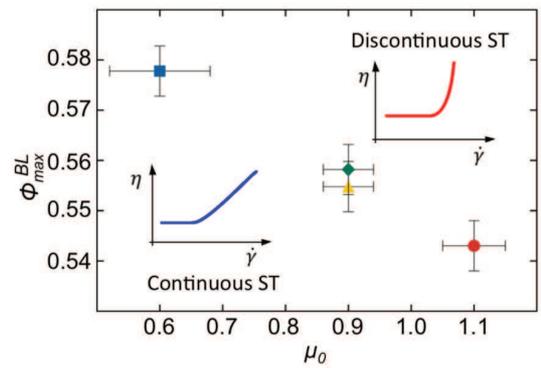}\\
\caption{(color online)$\phi^{BL}_{max}$ as a function of the coefficient of friction in boundary regime $\mu_0$ for the four polymers (same symbols as in Fig.\ref{rheo}). The CST and DST regions are highlighted in the graph. }\label{PhiVsMu}
\end{figure}

}


Using a simple theoretical framework, independently backed up by simulations and experiments, we have identified the microscopic origin of both continuous and discontinuous shear-thickening of dense non-Brownian suspensions as the consequence of the transition from hydrodynamically lubricated to boundary lubricated contacts. The central role played by friction introduces the local Sommerfeld number as the controlling parameter for the transition between Newtonian and shear-thickening regimes, as demonstrated by our numerical simulations. The presence of two distinct lubrication regimes as a function of the Sommerfeld number is furthermore at the origin of the Nw-ST transition. In particular, the friction coefficient in the boundary regime, which we tuned experimentally by polymer adsorption on the particle surface, governs the nature of the ST transition. Distinct lubrication regimes imply that the jamming volume fractions in the viscous regime $\phi^{HD}_{max}$ and in the Bagnoldian regime $\phi^{BL}_{max}$ are not the same in general, given that only the latter depends on the friction coefficient. Therefore CST is found when $\phi^{HD}_{max} \geq \phi^{BL}_{max} \geq \phi$, while the suspension exhibits DST when the transition to the inertial regime is impossible because $\phi^{HD}_{max} \geq \phi>\phi^{BL}_{max}$. Thus, in the absence of transient migration effects \cite{Fall:2010hu}, the local volume fraction and friction coefficient determine the stable microscopic flow mechanism, which is either CST or DST \cite{Fall:2008ud,Fall:2010hu,Fall:2012bv}. Moreover, our model does not require any confinement at the boundaries, but only that locally $\phi> \phi^{BL}_{max} $. This condition is fulfilled by preventing particle migration out of the shear zone, either by confinement during steady-state shear \cite{Brown:2010wt} or by keeping the shear duration short enough \cite{Waitukaitis:2012dv}.

The generality and consistency of our data and of the proposed model sets a global framework in which the tribological (friction) and rheological properties of dense non-colloidal systems are intimately connected. This concept is expected to have an impact on a host of practical applications and relates fundamental issues such as flow localization \cite{Huang:2005wv} and the solid-liquid-solid transition of granular pastes \cite{Fall:2008wh}.

\begin{acknowledgments}
$Acknowledgments$ - The authors thank Fabrice \textsc{Toussaint} for scientific discussions during preliminary work and C\'{e}dric \textsc{Juge}, Abdelaziz \textsc{Labyad} and Serge \textsc{Ghilardi} for technical support.

The authors acknowledge financial support by Lafarge LCR. Furthermore, this work was supported by the FP7-319968 grant of the European Research Council, the Ambizione grant PZ00P2$\_$142532/1 of the Swiss National Science Foundation and the HE 2732/11-1 grant of the German Research Foundation.

\end{acknowledgments}

\bibliography{biblio}

\end{document}